\documentclass[12pt]{iopart}

\usepackage{natbib}
\usepackage{overpic}
\usepackage{longtable}

\begin{document}


\title[Supernovae and the IMF in disturbed galaxies]{Type Ibc
  supernovae in disturbed galaxies: evidence for a top-heavy IMF}


\author{S M Habergham$^1$, J P Anderson$^2$ and P A James$^1$}
\address{$^1$ Astrophysics Research Institute, Liverpool John Moores
  University, Twelve Quays House, Birkenhead, CH41 1LD}
\address{$^2$ Departamento de Astronom\'ia, Universidad de Chile, Casilla
  36-D, Santiago, Chile}
\ead{smh@astro.livjm.ac.uk}


\begin{abstract}
We compare the radial locations of 178 core-collapse supernovae to the
$R$-band and H$\alpha$ light distributions of their host galaxies.
When the galaxies are split into `disturbed' and `undisturbed'
categories, a striking difference emerges. The disturbed galaxies
have a central excess of core-collapse supernovae, and this excess is
almost completely dominated by supernovae of types Ib, Ic and Ib/c,
whereas type II supernovae dominate in all other environments. The
difference cannot easily be explained by metallicity or extinction
effects, and thus we propose that this is direct evidence for a
stellar initial mass function that is strongly weighted towards high mass
stars, specifically in the central regions of disturbed galaxies.
\end{abstract}



\newpage



\section{Introduction}
\label{sec:intro}

Following the pioneering work of \citet{lars78}, many studies have
confirmed that tidal disturbance following galaxy interactions is an
efficient trigger of star formation in galaxies
\citep[e.g.][]{jose84,kenn84}. Such star formation frequently takes
the form of centrally-concentrated nuclear starbursts \citep{jose85},
fuelled by the central concentrations of molecular gas found to occur
naturally in simulations of highly-disturbed systems
\citep{barn91,miho96}. The strength of the link between starbursts
and interactions was highlighted by the finding that almost all of the
`ultra-luminous infrared galaxies' (ULIRGs) display signs of
interactions or mergers \citep{sand88,born99}, and by correlations
between galaxy-galaxy separations and starburst strength
\citep{bart00}. Even minor mergers with low-mass companions have been
shown through simulations to result in significant nuclear star
formation activity \citep{miho94}.

Several early studies of nuclear starbursts suggested that this star
formation might require a top heavy initial mass function (IMF),
preferentially producing high mass stars \citep{riek80, doyo92}.
There is theoretical support for this suggestion, with simulations
showing that an IMF weighted to high-mass stars naturally
arises in high-density regions, due to feedback processes heating
the gas. In a recent study, \citet{krum10} have
demonstrated that such regions should have a high-mass stellar
fraction at least 1.7 times larger, and possibly much more, than lower
density, more quiescent regions.

However, the observational evidence for this variation has to date
proved controversial (see \citealt{bast10} for a recent review). Some
studies have found indirect evidence for top-heavy IMFs with, for
example, \citet{riek93} concluding that the nearby starburst galaxy
M82 requires an IMF biased to high mass stars to explain its emission
line ratios and total luminosity. Similar techniques have been used
for NGC 3256 (an ongoing merger with a `super-starburst') which have
again shown indications of a modified IMF with an excess of high mass
stars \citep{doyo94}. \citet{gibs97} showed that, in order to
reproduce the observed colour-luminosity relation of elliptical
galaxies, an IMF much flatter than that of \citet{salp55} needed to be
adopted. \citet{baug05} had to employ a top heavy IMF for the
starbursts powering the distant population of highly luminous
submillimetre galaxies in order to explain the number counts of these
systems.  Finally, \citet{bras07} studied nine interacting galaxies
from the $Chandra$ survey and found that highly disturbed systems
showed a strongly enhanced infrared luminosity compared to that
expected from the x-ray emission, again suggesting the need for a
top-heavy IMF.

More direct evidence of a variation in IMF has been found for the
resolved stellar population of the young Arches cluster in the
Galactic Centre. \citet{fige99,stol02,paum06,espi09} all find
evidence for stellar mass functions weighted towards high-mass stars
in this cluster or the general Galactic Centre region. Such mass
functions are parametrized as an IMF that is either much flatter than that
found by \citet{salp55}, or having a higher mass turnover than is found in 
the function for field stars.
 
One possible tracer of the IMF that has not been fully exploited to
date is the relative numbers of core-collapse supernovae (CCSNe) of
different types. Their short progenitor lifetimes and high
luminosities make them powerful indicators of recent or ongoing star
formation, and indeed they provide the only direct tracer of recent
star formation within unresolved stellar populations. Recent advances
in the understanding of supernovae and their progenitors raise the
possibility that they can provide information on the initial mass
function of a young stellar population. Theoretical models of single
star progenitors predict that SNII should have lower mass progenitors
than SNIb or SNIc \citep{hege03, eldr04}. This has received
observational support from studies of the strength of association with
H$\alpha$ emission \citep{ande08}, confirming that SNII have the
lowest mass CC progenitors, but additionally indicating that the SNIc
have still higher mass progenitors than the SNIb. The existence of
this II-Ib-Ic progenitor mass sequence allows information on the IMF
of the stellar population in the SN environments to be derived from
the relative numbers of type II, Ib and Ic supernovae.

\citet{petr95} studied the distribution of SNe events in 32
interacting systems containing 12 known core collapse SNe. They found
that the radial distribution of these core collapse events showed a
higher concentration towards the nuclear regions of the interacting
galaxies when compared to isolated galaxies. This confirmed the
enhanced star formation around the central regions of the systems, but
the sample was too small to analyse the separate types of CCSNe.

This paper will therefore use a larger sample of local CCSNe to
explore the IMF in nuclear starbursts, resulting from galaxy
disturbance, by studying the ratio of type II/Ibc SNe in both
disturbed and undisturbed host galaxies. Throughout this paper, we
use `Ibc' to encompass all SNe with classifications of Ib, Ic or Ib/c.

The structure of the paper is as follows: In Section~\ref{sec:sample}
we will define and discuss the sample used throughout this
work. Section~\ref{sec:results} will describe the results on the
radial distributions, for disturbed and undisturbed hosts and looking
separately at type II and Ibc SNe. In Section~\ref{sec:disc} we
discuss the possible interpretations of our results, in terms of
metallicity, extinction and IMF effects. Finally,
Section~\ref{sec:conc} contains a summary of our conclusions.

\section{Sample and observations}
\label{sec:sample}

The sample used in this work consists of 140 local (recession
velocity $<$6000~km/s) spiral galaxies, hosts to 178 CCSNe (110
  SNII and 68 SNIbc), for which we have H$\alpha$ and $R$-band
observations from the Liverpool Telescope (LT) and Isaac Newton
Telescope (INT). (Some galaxies do not have usable images in
  either H$\alpha$ or $R$-band and have been omitted from the
  corresponding plots and statistics; see Tables 1 \& 2 in the online
  material).  This is the same dataset as was used by \citet{ander09}
with a small number of subsequent observations. SNe classified as type
IIb are not included in this sample as they are thought to be
transitional objects between SNII and SNIb, with substantially larger
progenitor masses ($\sim$25 M$_{\odot}$; \citealt{smar09}) than
typical SNII. A comparison performed on January 21st 2010 with all
CCSNe host galaxies within the same recession velocity limit in the
IAU SN
catalogue\footnote{http://www.cfa.harvard.edu/iau/lists/Supernovae.html},
and where the SNe have accurate classifications and positions, showed
this sample to be $\sim$34\% complete for SNIbc and $\sim$18\%
complete for SNII.

The classification of host galaxies as disturbed is purely by visual
inspection by the authors and thus is subjective. Galaxies which show
signs of tidal tails, definite interaction, double nuclei or strong
asymmetry have therefore been classed as disturbed.

\section{Results}
\label{sec:results}

The total sample of CCSNe is dominated by SNII (62\% of the
total). When the sample is constrained only to supernovae which lie in
disturbed hosts (64 CCSNe) this falls to 56\% SNII,
compared to 65\% SNII in the non-disturbed hosts.

For each of the CCSNe in our sample we have calculated the Fr($R$) and
Fr(H$\alpha$) statistics used, and explained fully, in
\citet{ander09}. Briefly, these represent the fractions of galaxy
emission, in the $R$-band and H$\alpha$ respectively, that lie within
the circle or ellipse which contains the SN. Thus Fr($R$)$=$0.0
corresponds to a supernova at the central $R$-band peak of the galaxy
emission, or closer to this peak than any H$\alpha$ emission, in the
case of Fr(H$\alpha$); whilst Fr$=$1.0 implies an extreme outlying
SN. If the emission is statistically a good tracer of the parent
population of supernovae, the Fr values should have a flat
distribution with a mean value of 0.5.

Figures 1 and 2 show the distributions of Fr($R$) values for the CCSNe
in the present sample, for the undisturbed and disturbed galaxies
respectively. In all histograms shown in this paper, the upper plot
represents the CCSNe sample, the middle the type II SNe and the lower
SNIbc. Looking first at the overall distributions of CCSNe, there is a
clear difference between the disturbed and undisturbed subsets, in the
sense that the disturbed galaxies have substantially more CCSNe
occurring in their central regions, with low Fr($R$) values. For
example, 36 of the 58 CCSNe in the disturbed sample occur
within the central 50\% of the $R$-band light, 62\% of the
total, compared with 50 out of 112 (45\%) in the undisturbed
galaxies. A Kolmogorov-Smirnov (KS) test shows that the chance of the
two total CCSNe distributions being drawn from the same parent
distribution is P$=$0.037. Thus there is evidence at the
2$\sigma$ level that galaxy disturbance correlates with
centrally-enhanced star formation and hence the production of an
increased central fraction of CCSNe.

The most striking aspect of Figure 2 is the types of SNe that make up
this central excess in the disturbed galaxies. Remarkably, given that
SNIbc only comprise 38\% of the overall CCSN sample (68/178), all 5 of
the CCSNe coming from the central 10\% of the disturbed host galaxy
light, and 11 of the 13 coming from the central 20\% of the light, are
of type Ibc. A KS test of the Fr($R$) distributions for the disturbed
galaxy subsample finds P$=$0.003, indicating a very low
probability that the SNIbc and SNII Fr($R$) values are drawn from the
same parent distribution. The mean values of Fr($R$) are 0.31 (95\%
confidence limits~ 0.20--0.42) for the SNIbc in the disturbed
galaxies, compared with 0.51 (0.44--0.59) for the SNII in
the disturbed galaxies. This is the main observational result from
this paper; the CCSNe occurring in the central regions of disturbed
galaxies are heavily weighted towards types Ib, Ic and Ib/c. We will
discuss possible interpretations of this in Section 4.

Some further statistical tests were also performed on the CCSN
distributions shown in Figures~1 \& 2. Figure~1 shows that even in the
undisturbed galaxies, there is some evidence for a larger fraction of
SNIbc in the central regions, principally due to a central `hole' in
the radial distribution of SNII. A KS test applied to the SNIbc and
SNII distributions shown in Figure 1 shows this difference to be only
marginal, P=0.082, and hence clearly less marked than for the
disturbed galaxies; disturbance does seem to play a part in the
central concentration of the SNIbc. This point was further explored
by comparing the SNIbc distributions for undisturbed and disturbed
galaxies, i.e. Figure 1 vs. Figure 2; this did indicate the SNIbc in
disturbed galaxies to be more centrally concentrated, with a KS P
value of 0.06, again of marginal significance. The mean SNIbc Fr($R$)
value is 0.48 (0.38--0.57) for the SNIbc in the undisturbed galaxies,
again to be compared with 0.31 (0.20--0.42) already quoted for the
disturbed galaxies. Finally for Figure 2, it might be asked whether
there is evidence for a {\it suppression} of SNIbc fraction in the
outer regions of these galaxies. However, given the current sample
size this cannot be determined with any significance. For example we
find 6 SNIbc in the outer 50\% of the light distributions of the
disturbed galaxies, but with only 22 CCSNe in total from these
regions, this is not significantly below the expectation value of 8.4,
based on the SNIbc/SNII ratio for the full sample.

Figures 3 \& 4 show the distributions of supernova locations relative
to the H$\alpha$ distributions of their host galaxies. Overall these
show the same patterns as Figures 1 \& 2, but they do enable one
specific issue to be addressed: are the SNIbc more centrally
concentrated than the H$\alpha$ light, which is presumably a good
tracer of the youngest stellar population? Figure 4 shows that there
is some evidence for this; the central 10\% of the H$\alpha$ emission
in the disturbed galaxies gives rise to 7 of the 22 SNIbc in these
galaxies. The mean Fr(H$\alpha$) value for the SNIbc in disturbed
galaxies is 0.33 (0.21--0.45), so this population does seem to be more
centrally concentrated than the H$\alpha$ emission. This is not true
for the SNIbc in the undisturbed galaxies, or for the SNII in either
of the galaxy subsets; all of these distributions have mean
Fr(H$\alpha$) values consistent with 0.5.

\section{Discussion}
\label{sec:disc}

\citet{ander09} found a central excess of SNIbc in a SN-host galaxy
sample. This work has found that this central excess is exaggerated in
galaxies which appear disturbed. A more centrally located distribution
of SNIbc has been suggested previously \citep[e.g.][]{bart92, petr95,
 vand97}, though previous studies often suffered from low number
statistics. \citet{hako09} also find SNIbc to be more centrally
located than SNII, however in conflict to our results they do not find
the central excess of SNIbc clearly seen in our data. An important
difference between our work and most other studies in the literature
is that our method implicitly normalizes the tests to the measured
distributions of different stellar populations; other studies use
distances normalized to isophotal radii. Most of these results have
been interpreted as an increase in metallicity of the SNIbc
progenitors, although \citet{hako09} also make the suggestion of a
shallower IMF within the central regions.

Studies conducted into active and star-forming galaxies \citep{petr05}
and Seyfert galaxies \citep{bres02} have also noted marginal evidence
for an increased fraction of both CCSNe and specifically SNIbc within
these galaxies when compared to `normal' ones.

There are various observational biases which may affect our
analysis. \citet{shaw79} found a bias in supernova samples, in the
sense that it is more difficult to detect SNe in the inner regions of
distant galaxies. The sample is also subject to any bias contained
within the object selection found in the Asiago \citep{barb09} and IAU
SN catalogues. For the Asiago and Crimea searches, \citet {capp93}
estimated the number of SNe lost due to overexposure combined with the
Shaw effect, which for the velocity range of our sample is
$\sim$35\%. Another source of bias is the loss of SNe in the central
regions of galaxies through the large amount of dust obscuration which
has been investigated through near infrared studies
\citep[e.g.][]{matt07, kank08}. Such biases should affect all SN
types, although the intrinsically fainter SNIIP \citep{rich02, rich06}
may be rather more likely to be lost through these effects.~However,
if our results are correct and SNIbc are more centrally concentrated
than SNII then recovering all of the lost central SNe would lead to an
even more exaggerated excess.

One possible source of error is our eyeball classifications of host
galaxy disturbance. In future we plan to quantify this through
near-IR observations and use of objective measures of asymmetry
\citep{cons00,lotz04}. However, we are quite confident in our
disturbance classifications; images of 12 of our `disturbed' galaxies
with centrally-located SNe are shown in Figure~5, confirming that this
is not a `normal' group of galaxies.

The high central excess of SNIbc in the central regions of the
disturbed host galaxies found in this work is difficult to explain in
terms of effects other than an IMF biased towards high-mass stars.~A
possible alternative explanation is the effect of metallicity, given
that \citet{bois09} find that the ratio of SNIbc/SNII increases with
both local and global metallicity. Looking at the absolute magnitudes
of the host galaxies, we do indeed find that the disturbed galaxies
are somewhat more luminous than the undisturbed galaxies (KS
probability of 0.07 that they are drawn from the same parent
distribution), by almost 0.4~mag in the mean, which might imply a
somewhat higher mean metallicity in the disturbed galaxies.  However,
this does not seem to be driving the result we find.  Splitting the
total sample (disturbed and undisturbed) by absolute magnitude, we
find no significant differences in the Fr($R$) distributions of bright
and faint galaxies.  Splitting into bright and faint halves, the KS
probability is 0.998 (complete consistency), whereas when the bright
third is compared with the faintest two-thirds (to better match the
disturbed/undisturbed split), P is 0.276, but in the sense that the
bright galaxies have a slight bias towards {\em high} Fr($R$) values.
In any case, the expected metallicity bias resulting from a difference
of 0.4 in galaxy absolute magnitude is very small.  The
mass-metallicity relation of \citet{trem04} for galaxies of a few
times 10$^{10}$~M$_{\odot}$ predicts a corresponding change of only
$\sim$0.025 dex in log(O/H), highly unlikely to cause any significant
effects. Finally, in interacting systems the central metallicity is
lowered by the in-fall of unenriched gas \citep{mich08, elli08,
  rupk10}. This would therefore act in the opposite sense to the
result we find. It should also be noted that whilst a study of
gas-phase metallicities of the local environments of CCSNe (Anderson
et al. MNRAS submitted) finds a trend favouring SNIbc in
high-metallicity regions, even the highest metallicity environments
seem to host a significant fraction of SNII.

It is also possible that stellar rotation \citep[e.g.][]{hege03} and
binarity \citep[e.g.][]{nomo95} could contribute to this effect. It is
not clear why the binary fraction should be higher within the
disturbed galaxy sample, but it should be noted that the increased
densities within these nuclear starburst regions could lead to more
massive and denser clusters, within which processes such as stellar
mergers and binary interactions would be more prevalent
\citep[e.g.][]{port10}.

To conclude, our preferred explanation of this central excess of SNIbc
is that the central regions of these disturbed galaxies are hosting
starbursts with initial mass functions biased to high stellar
masses. Given the small numbers of SNe involved, the uncertain mass
limits corresponding to progenitors of different SN types, and the
likely role of binarity in determining SN type, it is hard to quantify
the implications of this result for the IMF. However, an illustrative
calculation can be performed as follows. Under the assumption of a
Salpeter IMF, and the (admittedly simplistic) assumptions that CCSNe
arise from single stars with masses between 8 and 80 M$_\odot$ and
that mass alone determines SN type, the relative numbers of SNII and
Ibc in the outer regions of undisturbed galaxies (2.3:1) indicate a
transition at about 18~M$_\odot$. The apparent inversion of this ratio
for the central regions of the disturbed galaxies (0.18:1), if
interpreted purely as a change in IMF slope, appears to require a
positive index in the IMF slope (formally $x=$+0.95 cf. --1.35,
assuming the transition mass is unchanged at 18~M$_\odot$). However, we
emphasize that this is purely illustrative; all of the assumptions are
likely to be in error at some level, and binarity and metallicity
effects may play some part in the changes we find.

We note here that we are modifying the conclusions of \citet{ander09}
in that it is hard to interpret the previously found central SNIbc
excess purely in terms of metallicity effects. However, we note that
there is still a marginal central excess of SNIbc in the undisturbed
galaxy sample, indicating some effect of metallicity. Quantifying the
relative sizes of the different effects will be the focus of future
studies.

Finally, it is interesting to note that with current research
indicating a connection between gamma ray bursts (GRBs) and type Ic
SNe \citep{woos06}, recent studies \citep[e.g.][]{cons05, wain07,
  frye07} have found that GRB host galaxies show an over-abundance of
merging or interacting galaxies compared to other star-forming hosts.

\section{Conclusions}
\label{sec:conc}

We have analysed the spatial distribution of 178 CCSNe within a sample 
of host galaxies with recession velocities less than 6000~km/s. Host 
galaxies were classified by eye according to whether they show disturbance
due to strong tidal interactions or mergers. The main results are as follows:

\begin{itemize}
\item CCSNe of all types show a strong degree of central concentration
  in the disturbed galaxies, probably as a result of nuclear
  starbursts in these galaxies.
\item This central excess is dominated by SNIbc.
\item The SNIbc in disturbed galaxies are more centrally concentrated than the
H$\alpha$ emission.
\item The SNIbc excess cannot easily be explained in terms of metallicity 
effects, extinction, or central incompleteness of SNe.
\item Our preferred explanation of the SNIbc excess is that the
  central regions of the disturbed galaxies are dominated by nuclear
  starbursts with IMFs biased towards high mass stars, although
  metallicity, binarity and stellar rotation may also play a role.
\end{itemize}

\ack

This paper has made use of data provided by the Central Bureau for
Astronomical Telegrams. We would like to acknowledge members of staff
at the Astrophysics Research Institute, in particular Sue Percival and
David Bersier for their helpful comments and discussion. The authors
would also like to thank the referee for a constructive and helpful
report. This research has made use of the NASA/IPAC Extragalactic
Database (NED) which is operated by the Jet Propulsion Laboratory,
California Institute of Technology, under contract with the National
Aeronautics and Space Administration. The Isaac Newton Telescope is
operated on the island of La Palma by the Isaac Newton Group in the
Spanish Observatorio del Roque de los Muchachos of the Instituto de
Astrof\'isica de Canarias. The Liverpool Telescope is operated on the
island of La Palma by Liverpool John Moores University in the Spanish
Observatorio del Roque de los Muchachos of the Instituto de
Astrof\'isica de Canarias with financial support from the UK Science
and Technology Facilities Council. SMH would like to acknowledge STFC
for a research studentship.

\newpage

\begin{figure}
\noindent\makebox[\textwidth]{%
\includegraphics{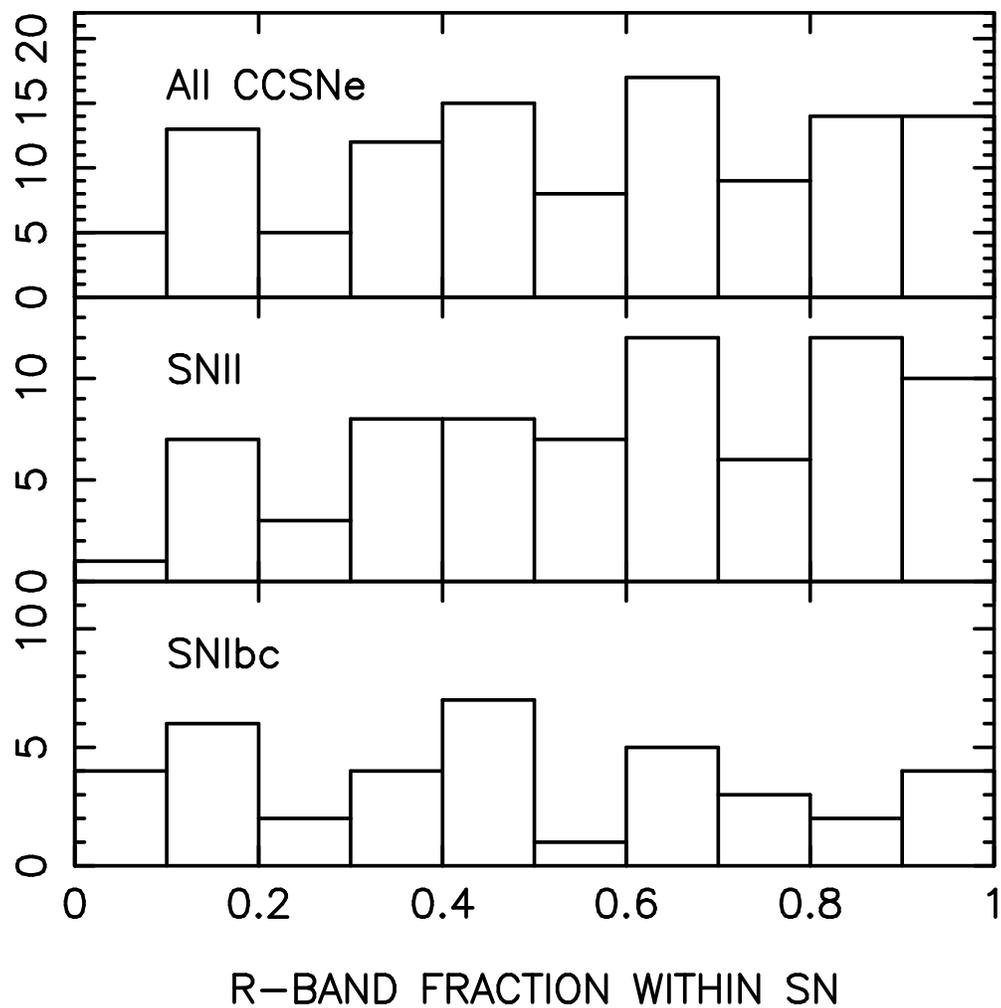}}
\caption{
 Histogram showing the distribution of fractions of host
 galaxy $R$-band light lying within the locations of each CCSN in our
 undisturbed host galaxies. The top plot represents the distribution
 of all CCSNe, the middle SNII and the lower SNIbc.}
\label{figure1}
\end{figure}

\newpage

\begin{figure}
\noindent\makebox[\textwidth]{%
\includegraphics{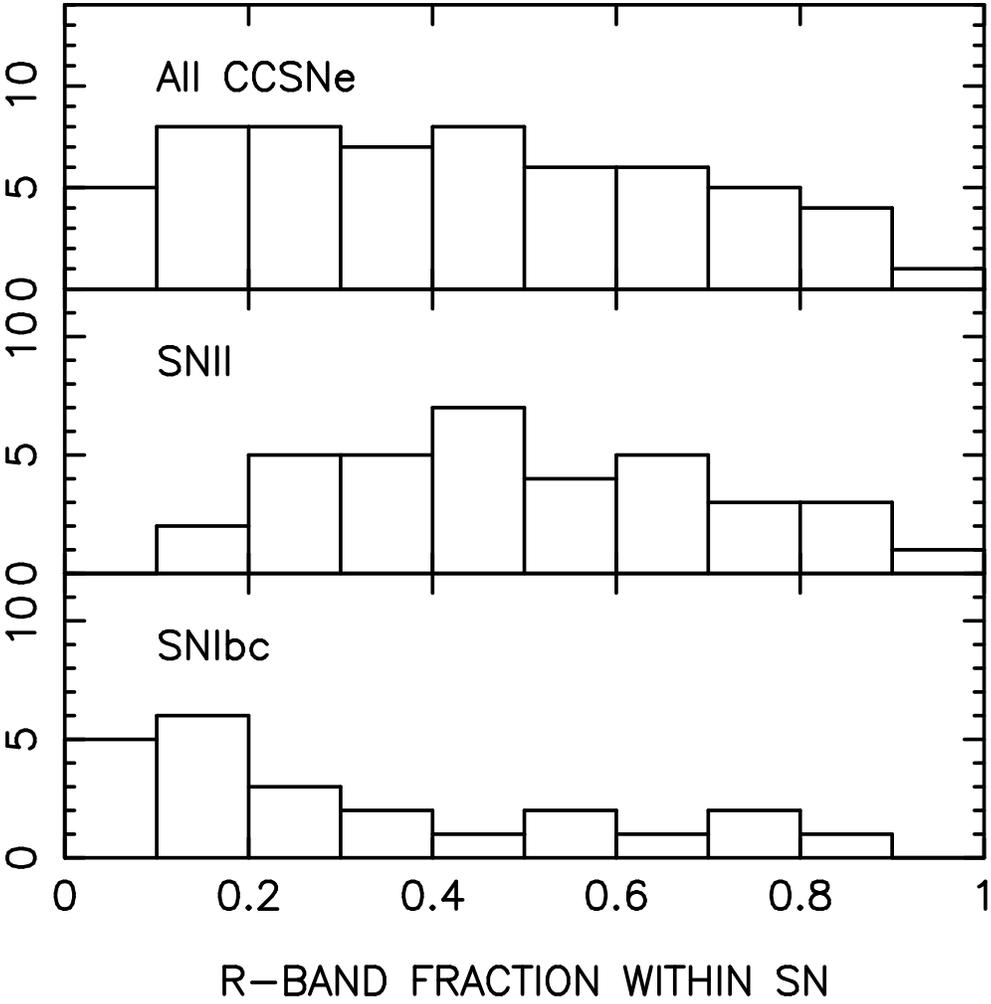}}
\caption{
As Figure 1, but for the disturbed host galaxies. 
 }
\label{figure2}
\end{figure}

\newpage

\begin{figure}
\noindent\makebox[\textwidth]{%
\includegraphics{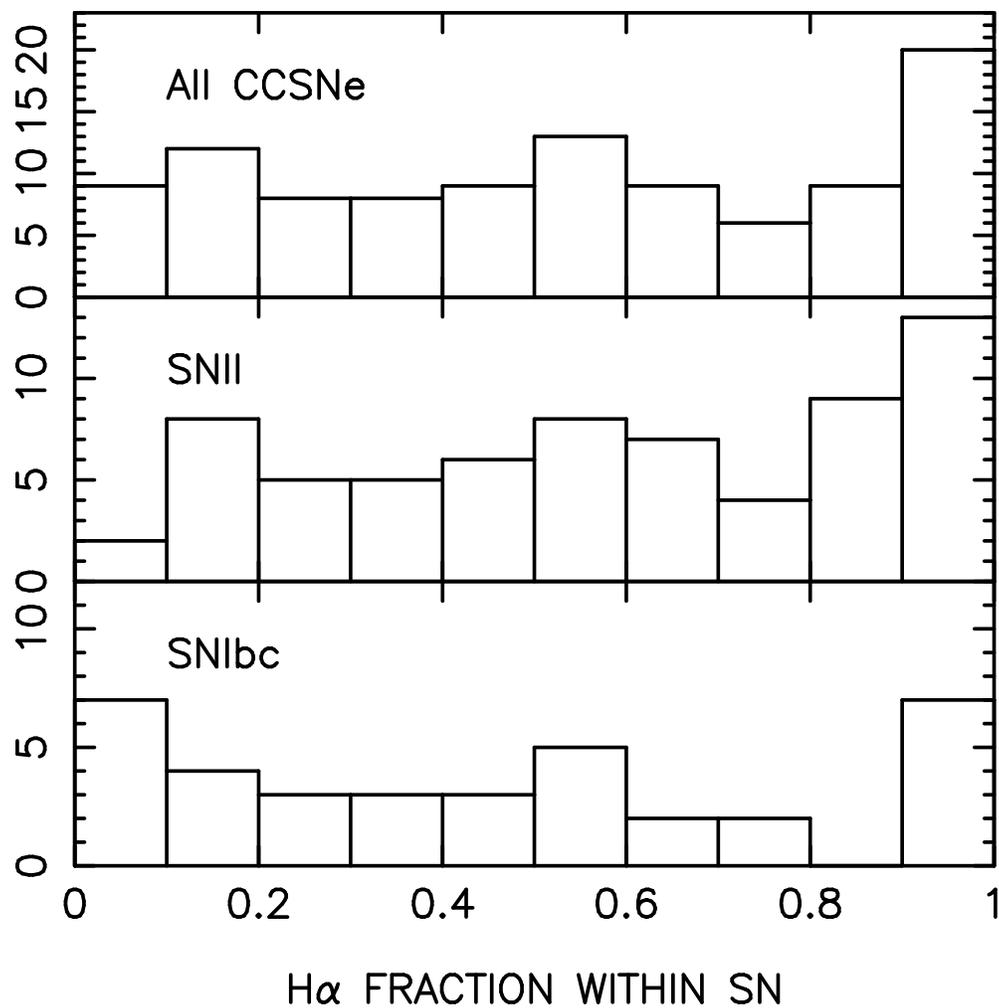}}
\caption{
Histogram showing the distribution of fractions of host galaxy
H$\alpha$ light lying within the locations of each CCSN in our sample,
for the undisturbed host galaxies. Again, the upper plot shows the
overall CCSNe distribution, the middle SNII and the lower SNIbc. 
 }
\label{figure3}
\end{figure}

\newpage

\begin{figure}
\noindent\makebox[\textwidth]{%
\includegraphics{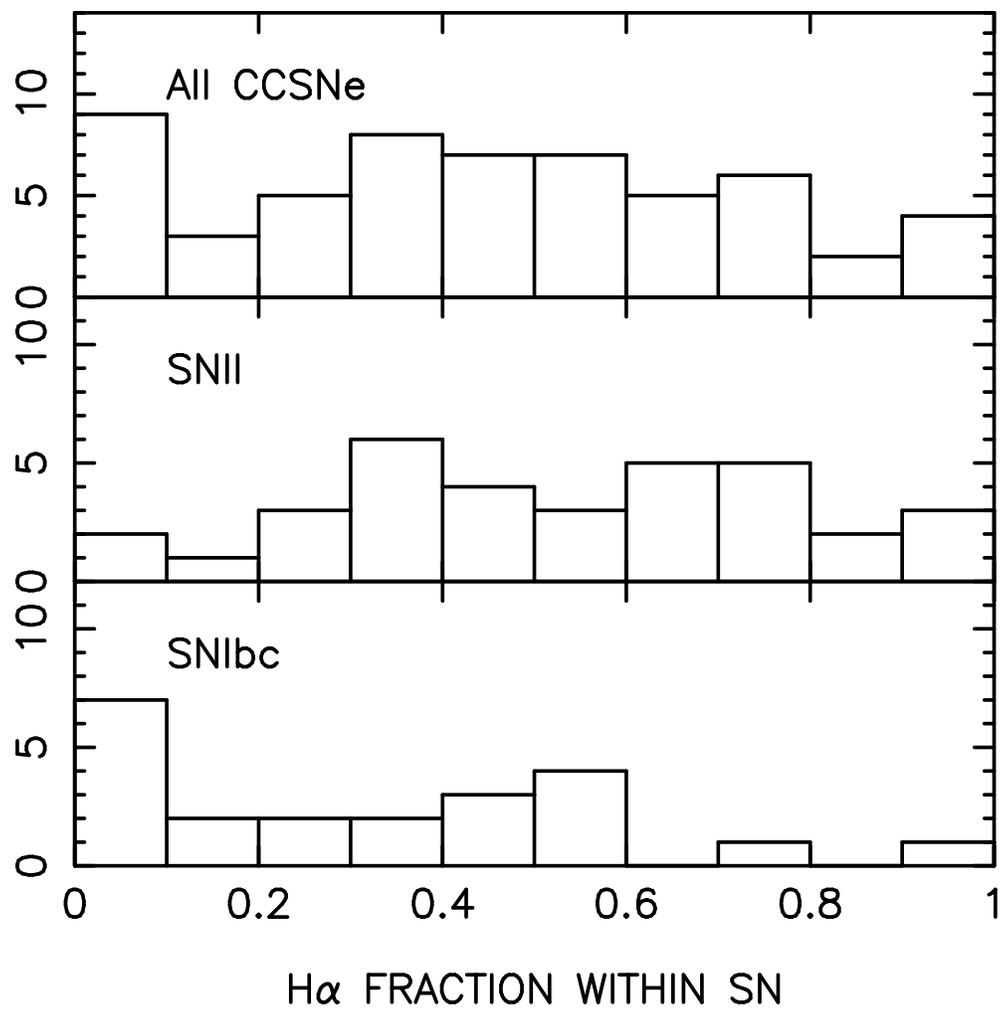}}
\caption{ 
As Figure 3, for the disturbed host galaxies. 
 }
\label{figure4}
\end{figure}

\newpage

\begin{figure}
\noindent\makebox[\textwidth]{%
\includegraphics{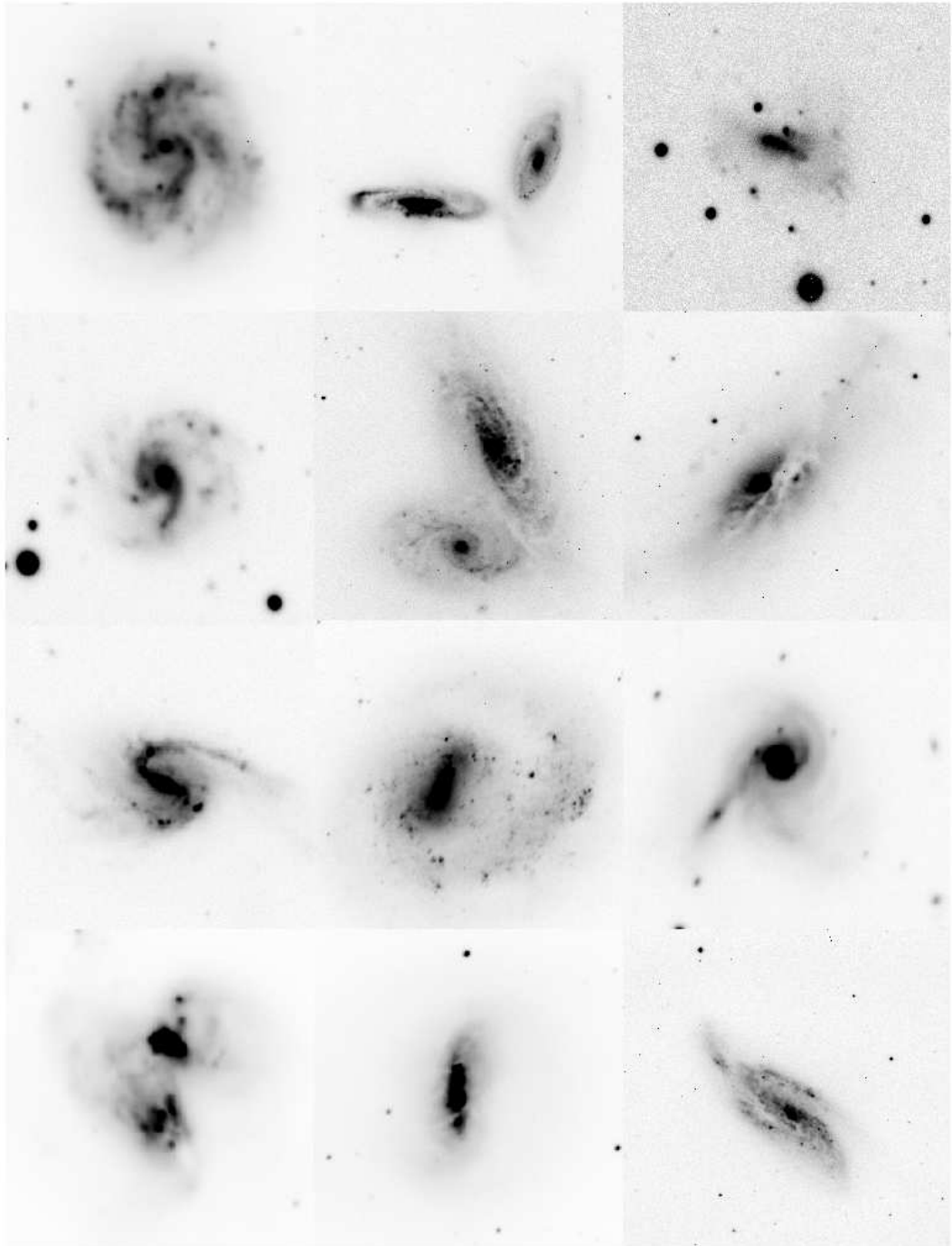}}
\caption{ 
Images of 12 of the host galaxies classified as disturbed and with
centrally located CCSNe. 
 }

\label{figure5}
\end{figure}

\newpage
\appendix
\setcounter{section}{1}

\Tables
\begin{indented}
\item
\begin{longtable}{lllll}
\caption{{\label{Table 1}}Undisturbed host galaxy sample used in this
  analysis. Columns represent the host galaxy, the individual SNe, the
  spectral classification of the SNe and the fractional $R$-band light
  and fractional H$\alpha$ values for each SNe.}\\ 
\br
Host&SN&SN type&Fr($R$)&Fr(H$\alpha$)\\ 
\mr
NGC493&1971S&IIP&0.605&0.570\\	
NGC918&2009js&IIP&0.703&-\\	
NGC941&2005ad&II&0.831&0.864\\
NGC991&1984L&Ib&0.498&0.401\\	
NGC1035&1990E&IIP&0.272&0.363\\
NGC1058&1961V&II&0.968&0.931\\	
NGC1058&1969L&IIP&1.000&1.000\\
NGC1058&2007gr&Ib/c&0.421&-\\	
NGC1073&1962L&Ic&0.754&0.518\\
NGC1087&1995V&II&0.368&0.497\\
NGC1187&1982R&Ib&0.695&0.760\\
NGC1187&2007Y&Ib&0.981&1.000\\	
MCG-01-09-24&2002ei&IIP&0.195&0.195\\
NGC1343&2008dv&Ic&0.195&0.134\\
UGC2906&2008im&Ib&0.682&-\\	
UGC2971&2003ig&Ic&0.176&0.108\\
IC381&2001ef&Ic&0.082&0.052\\	
NGC1832&2004gq&Ib&0.672&0.328\\
NGC1832&2009kr&II&0.489&-\\
IC2152&2004ep&II&0.461&0.560\\
UGC3804&2002A&IIn&0.419&0.253\\
NGC2551&2003hr&II&0.914&1.000\\	
NGC2596&2003bp&Ib&0.486&0.362\\
UGC4436&2004ak&II&0.887&0.882\\
NGC2726&1995F&Ic&0.037&0.050\\
NGC2742&2003Z&IIP&0.675&0.736\\
NGC2715&1987M&Ic&0.129&0.044\\
UGC4904&2006jc&Ib/c&0.332&0.525\\
NGC2841&1972R&Ib&0.855&0.904\\
NGC2906&2005ip&II&0.399&0.528\\
UGC5249&1989C&IIP&0.017&0.058\\
NGC3074&1965N&IIP&0.110&0.059\\
NGC3074&2002cp&Ib/c&0.936&0.961\\
NGC3147&2006gi&Ib&0.984&0.991\\
NGC3184&1921B&II&0.856&0.954\\
NGC3184&1937F&IIP&0.808&0.930\\
NGC3184&1999gi&IIP&0.276&0.112\\
NGC3198&1966J&Ib&0.898&0.963\\
NGC3198&1999bw&IIn&0.745&0.755\\
NGC3240&2001M&Ic&0.323&0.251\\
NGC3294&1990H&IIP&0.156&0.125\\
NGC3340&2005O&Ib&0.322&0.305\\
\newpage
\br
Host&SN&SN type&Fr($R$)&Fr(H$\alpha$)\\ 
\mr
NGC3340&2007fp&II&0.170&0.125\\
NGC3430&2004ez&II&0.788&0.833\\
NGC3437&2004bm&Ic&0.073&0.076\\
NGC3451&1997dn&II&0.872&0.946\\
NGC3504&2001ac&IIn&0.826&0.992\\
NGC3512&2001fv&IIP&0.669&0.689\\
NGC3556&1969B&IIP&0.197&0.494\\ 
NGC3631&1964A&II&0.915&0.992\\ 
NGC3631&1965L&IIP&0.622&0.658\\
NGC3631&1996bu&IIn&0.923&0.993\\
NGC3655&2002ji&Ib/c&0.709&0.957\\
NGC3683&2004C&Ic&0.532&0.545\\
UGC6517&2006lv&Ib/c&0.480&-\\
NGC3756&1975T&IIP&0.846&0.856\\
NGC3810&2000ew&Ic&0.261&0.147\\
NGC3810&1997dq&Ib/c&0.774&0.734\\
NGC3949&2000db&II&0.364&0.253\\
NGC3963&1997ei&Ic&0.197&0.053\\
NGC4030&2007aa&II&0.942&0.828\\
NGC4041&1994W&IIn&0.491&0.541\\
NGC4051&1983I&Ic&0.498&0.473\\
NGC4051&2003ie&II&0.838&0.885\\
IC758&1999bg&IIP&0.669&0.657\\
NGC4136&1941C&II&0.880&0.882\\
NGC4210&2002ho&Ic&0.146&0.051\\
NGC4242&2002bu&IIn&0.896&0.930\\
NGC4303&1926A&IIL&0.607&0.736\\
NGC4303&1961I&II&0.697&0.877\\
NGC4303&1964F&II&0.189&0.106\\
NGC4303&1999gn&IIP&0.418&0.429\\
NGC4303&2006ov&IIP&0.418&0.429\\
NGC4303&2008in&IIP&0.845&-\\
NGC4369&2005kl&Ic&0.271&0.540\\
NGC4384&2000de&Ib&0.087&0.140\\
NGC4451&1985G&IIP&0.138&0.212\\
NGC4559&1941A&IIL&0.208&0.131\\
UGC7848&2006bv&IIn&0.579&-\\
NGC4666&1965H&IIP&0.324&0.198\\
NGC4708&2003ef&II&0.335&0.352\\
NGC4725&1940B&IIP&0.675&0.802\\
NGC4900&1999br&IIP&0.786&0.932\\
NGC4961&2005az&Ic&0.426&-\\
NGC4981&2007C&Ib&0.320&0.236\\
NGC5012&1997eg&IIn&0.503&0.449\\
NGC5033&1950C&Ib/c&0.972&1.000\\
NGC5033&1985L&IIL&0.585&0.571\\
NGC5334&2003gm&IIn&0.480&-\\	
NGC5371&1994Y&IIn&0.355&0.212\\
NGC5468&2002ed&IIP&0.811&0.791\\
\newpage
\br
Host&SN&SN type&Fr($R$)&Fr(H$\alpha$)\\ 
\mr
NGC5559&2001co&Ib/c&0.618&0.497\\
NGC5584&1996aq&Ic&0.178&0.086\\
NGC5630&2005dp&II&0.534&0.590\\
NGC5630&2006am&IIn&0.604&0.617\\
NGC5673&1996cc&II&0.924&0.934\\
NGC5668&2004G&II&0.657&0.595\\
NGC5775&1996ae&IIn&0.757&0.671\\
NGC5806&2004dg&IIP&0.484&0.378\\
NGC5850&1987B&IIn&0.995&-\\ 
NGC5879&1954C&II&0.615&0.511\\ 
NGC5921&2001X&IIP&0.579&0.369\\
NGC6118&2004dk&Ib&0.673&0.626\\
NGC6207&2004A&IIP&0.729&0.660\\
UGC10862&2004ao&Ib&-&0.215\\
NGC6643&2008ij&IIP&0.519&0.620\\
NGC6643&2008bo&Ib&0.451&0.510\\
NGC6700&2002cw&Ib&-&0.642\\
NGC6946&2004et&II&0.975&-\\
NGC6951&1999el&IIn&0.320&0.259\\
UGC11861&1995ag&II&0.343&0.170\\
UGC11861&1997db&II&0.633&0.396\\
UGC12160&1995X&II&0.564&0.484\\
UGC12182&2006fp&IIn&1.000&1.000\\
\br
\end{longtable}
\end{indented}
\newpage
\begin{indented}
\item
\begin{longtable}{lllll}
\caption{\label{Table 2}Disturbed host galaxy sample used in this
  analysis. Columns represent the host galaxy, the individual SNe, the
  spectral classification of the SNe and the fractional $R$-band light
  and fractional H$\alpha$ values for each SNe, as in table 1.}\\
\br
Host&SN&SN type&Fr($R$)&Fr(H$\alpha$)\\
\mr
NGC895&2003id&Ic&0.524&-\\
UGC2984&2002jz&Ic&0.091&0.099\\	
NGC1614&1996D&Ic&0.275&-\\
NGC1637&1999em&IIP&0.276&0.268\\
IC391&2001B&Ib&0.062&0.060\\
NGC1961&2001is&Ib&-&0.749\\
NGC2207&1999ec&Ib&-&0.521\\
NGC2207&2003H&Ib&-&0.259\\
NGC2146&2005V&Ib/c&0.033&0.091\\
ESO492-G2&2005lr&Ic&-&0.005\\
UGC3829&2001ej&Ib&0.152&0.391\\
NGC2276&1968V&II&0.699&0.790\\
NGC2276&2005dl&II&0.247&0.099\\
NGC2276&1993X&II&0.899&0.619\\
NGC2532&1999gb&IIn&0.485&0.443\\
NGC2532&2002hn&Ic&0.023&0.011\\
NGC2604&2002ce&II&0.381&0.560\\
NGC2782&1994ak&IIn&0.725&0.977\\
NGC2993&2003ao&IIP&0.456&0.784\\
NGC3169&1984E&IIL&0.684&0.731\\
NGC3310&1991N&Ic&0.268&0.277\\
NGC3323&2004bs&Ib&0.191&0.119\\
NGC3323&2005kk&II&0.766&0.875\\
NGC3367&1992C&II&0.689&0.687\\
NGC3367&2007am&II&0.302&0.314\\
NGC3627&1973R&IIP&0.471&0.566\\
NGC3627&1997bs&IIn&0.362&0.348\\
NGC3627&2009hd&II&0.496&-\\
NGC3690&1993G&IIL&0.464&0.744\\
NGC3690&1998T&Ib&0.056&0.056\\
NGC3690&1999D&II&0.560&0.849\\
NGC3786&1999bu&Ic&0.180&0.522\\
NGC3811&1971K&IIP&0.809&0.900\\
IC2973&1991A&Ic&0.742&0.588\\
NGC4038&2004gt&Ib/c&0.834&0.991\\
NGC4088&1991G&IIP&0.466&0.453\\
NGC4088&2009dd&II&0.100&-\\
NGC4141&2008X&IIP&0.194&0.085\\
NGC4141&2009E&IIP&0.594&0.491\\
NGC4254&1967H&II&0.664&0.648\\
NGC4254&1972Q&IIP&0.811&0.791\\
NGC4254&1986I&IIP&0.334&0.318\\
NGC4273&2008N&IIP&0.300&0.316\\
NGC4273&1936A&IIP&0.569&0.598\\
NGC4490&1982F&IIP&0.277&0.202\\
\newpage

\br
Host&SN&SN type&Fr($R$)&Fr(H$\alpha$)\\
\mr
NGC4568&1990B&Ic&0.302&-\\
NGC4568&2004cc&Ic&0.158&-\\
NGC4618&1985F&Ib&0.121&0.087\\
NGC4615&1987F&IIn&0.489&0.333\\
NGC4688&1966B&IIL&0.571&0.454\\
NGC4691&1997X&Ic&0.171&0.472\\
NGC5000&2003el&Ic&0.482&0.476\\
NGC5021&1996ak&II&0.619&0.659\\
MCG-04-32-07&2003am&II&0.211&0.339\\
NGC5194&1994I&Ic&-&0.122\\
NGC5194&2005cs&IIP&-&0.222\\
NGC5395&2000cr&Ic&0.538&0.549\\
NGC5480&1988L&Ib&0.230&0.369\\
NGC5682&2005ci&II&0.204&0.191\\
NGC7479&1990U&Ic&0.603&0.488\\
NGC7479&2009jf&Ib&0.764&-\\
NGC7537&2002gd&II&0.759&0.685\\
NGC7714&2007fo&Ib&0.377&-\\
UGC12846&2007od&IIP&0.945&1.000\\

\br
\end{longtable}
\end{indented}

\clearpage
\def\newblock{\hskip .11em plus .33em minus .07em}

\end{document}